\documentclass[a4paper,10pt,twocolumn,journal,final]{IEEEtran}

\IEEEoverridecommandlockouts

\usepackage{epsfig}
\usepackage{amsmath}
\usepackage{amssymb}
\usepackage{psfrag}
\usepackage[T1]{fontenc}
\usepackage{graphicx}

\graphicspath{{../figure/}}

\newtheorem{define}{Definition}
\newtheorem{prop}[define]{Proposition}

\newtheorem{theo}[define]{Theorem}

\newtheorem{lem}[define]{Lemma}

\newcommand{\eps}{\varepsilon}
\newcommand{\w}{\omega}
\newcommand{\Z}{\mathbb{Z}}

\title{\LARGE \bf Topological identification in networks of dynamical systems}

\author{
	Donatello Materassi\IEEEauthorrefmark{1} and
	Giacomo Innocenti\IEEEauthorrefmark{7} \medskip\\
	\IEEEauthorrefmark{1}
		Department of Electrical and Computer Engineering,\\
		University of Minnesota,\\
		200 Union St SE, 55455, Minneapolis (MN) \\
		{\tt\small mater013@umn.edu} \medskip\\
	\IEEEauthorrefmark{7}
		Dipartimento di Sistemi e Informatica,\\
		Universit\`a di Firenze,\\
		via di S. Marta 3, 50139 Firenze, Italy \\
		{\tt\small giacomo.innocenti@gmail.com}
}

\begin{document}

\maketitle
\thispagestyle{empty}
\pagestyle{empty}

\begin{abstract}
The paper deals with the problem of reconstructing the topological structure
of a network of dynamical systems.
A distance function is defined in order to evaluate the ``closeness'' of
two processes and a few useful mathematical properties are derived.
Theoretical results to guarantee the correctness of the identification procedure for networked linear systems with tree topology are provided as well.
Finally, the application of the techniques to the analysis of an actual complex network, i.e. to high frequency time series of the stock market, is illustrated.
\end{abstract}

\section{Introduction}
Under the influence of improved numerical tools, a significant interest for complex systems has been shown in many scientific fields.
In particular, attention has been focused on networks, highlighting the emergence of complicated phenomena from the connection of simple models.
To this regard, a relevant impulse has been provided by the advances in neural network theory, that has contributed to underline the importance of the connection topology in the realization of complex dynamics \cite{rojas96}.
As a consequence, graph theory \cite{dies06} has been successfully exploited to perform novel modeling approaches in several fields, such as Economics (see e.g. \cite{mant99,ms00,nrm07}), Biology (see e.g. \cite{eisen1998clu,rava02}) and Ecology (see e.g. \cite{bib:bunn00,bib:monestiez05,bib:urban01}), especially when the investigated phenomena were characterized by spatial distribution and a multivariate analysis technique is preferred \cite{gd08, gdIFAC08}.\\
To the best knowledge of the authors, there are very few theoretical results about the reconstruction of an unknown topology from data.
In this paper, we will focus our attention on tree topology networks.
Though its reduced complexity with respect to cyclic link structures, the tree connection model turns out to be particularly suitable to represent a large variety of processes.
In particular, the tree network scheme results effective in the description of systems with transportation, such as water and power supply, air and rail traffic, vascular systems of living organisms and channel and drainage networks (see e.g. \cite{bib:banavar00,bib:monestiez05,bib:bailly06,bib:durand07,bib:shao07}).
It is worth to highlight that this kind of models is deeply related to the idea of delay, that characterizes the connections as transportation media.
It is also important to recall that in linear dynamical system theory the transfer function is a powerful representation tool for delayed processes \cite{ljung99,ksh00}.\\
In many situations, when the topology to be reconstructed is a tree, the only observable nodes are the leaves. Then, the usual theoretical framework is almost always set in standard graph theory as in the Unweighted Pair Group Method with Arithmetic mean (UPGMA) \cite{MicSok57}. Its application is mainly in the reconstruction of evolutionary trees, but it has been widely employed also in many other areas: communication systems and resource allocations.
Theoretically, such a technique guarantees an exact reconstruction of a tree topology only on the strong assumption that an ultrametric is defined among the considered nodes.
An approach based on system theory and identification tools is completely missing.
Specifically, there are no approaches considering explicitly the possibility of dynamics among the nodes. While dynamical networks have been deeply studied and analyzed in automatic control theory, the question of reconstructing an unknown network of dynamical systems has not been formally investigated. In fact, in most applicative scenarios the network is given or it is the very objective of design. However, there are also some interesting situations where the network links are actually unknown, such as in biological neural networks, biochemical metabolic pathways and financial markets.
Even though an acyclical topology may seem a quite reductive choice, given an intricate and connected topology, we may be interested into ``approximating'' it with a tree. Such an approximation could be considered ``satisfactory'' if the most important connections were captured.\\
In this manuscript we will develop a rigorous mathematical method to exactly identify the connections scheme of a tree topology network of noisy linear dynamical systems, providing a theoretical background for linear network modeling.
In particular, in Section \ref{sec:prob setup} we will introduce definitions and preliminary results which are useful to characterize the mathematical framework.
In Section \ref{sec:theory} our approach to topology reconstruction will be presented and sufficient conditions for an exact identification will be reported as well.
In Section \ref{sec:numeric} the theoretical results will be confirmed by practical implementations of the proposed technique, illustrated by means of numerical examples.
In Section \ref{sec:portfolio}, we will show that the identification of a tree topology can provide useful information even for complex network.
To this end, we will apply our technique to the analysis of high frequency real data originated by a portfolio of financial stocks.
Some final conclusions in Section \ref{sec:conclusions} will end the manuscript.\\
~\\
\noindent{\bf Notation:}\\
$E[\cdot]$: mean operator; \\
$R_{XY}(\tau)\doteq E[X(t)Y(t+\tau)]$: cross-covariance function of stationary processes;\\
$R_{X}(\tau)\doteq R_{XX}(\tau)$: autocovariance;\\
$\rho_{XY}\doteq\frac{R_{XY}}{\sqrt{R_X R_Y}}$: correlation index;\\
$\mathcal{Z}(\cdot)$: Zeta-transform of a signal;\\
$\Phi_{XY}(z)\doteq\mathcal{Z}(R_{XY}(\tau))$: cross-power spectral density;\\
$\Phi_{X}(z)\doteq\Phi_{XX}(z)$: power spectral density;\\
with abuse of notation, $\Phi_{X}(\w)=\Phi_{X}(e^{i\w})$; \\
$\lceil \cdot \rceil$ and $\lfloor \cdot \rfloor$:
			ceiling and floor function respectively;\\
$(\cdot)^*$: complex conjugate.

\section{Problem set up}\label{sec:prob setup}

In this section we formally introduce a model to address noisy linear dynamical systems interconnected to form a tree topology and we also provide a quantitative tool to characterize the mutual dependencies.

Let us consider a network of $n$ time-discrete SISO linear dynamical systems affected by additive noises.
Then, let $H_j(z)$ be the transfer function of the $j$-th system, $\{X_j(k)\}_{k\in\Z}$
and $\{U_j(k)\}_{k\in\Z}$ its output and input signals respectively and $\{\varrho_j(k)\}_{k\in\Z}$ a zero-mean wide-sense stationary noise.
Hence, each system can be represented according to the model:
\begin{align}
\label{eq:noisy sys}
	X_j(k) = H_j(z)U_j(k)+\varrho_j(k) \quad \forall j=1,\ldots,n \,.
\end{align}
We stress that no assumptions on the causality of $H_j(z)$ have been done.
Moreover, let the property
\begin{align}
\label{eq:uncorrelation}
	E[\varrho_j\varrho_i] = 0 \quad \forall j\neq i \,,
\end{align}
holds.
Then, suppose that the input signal $U_i$ of each node results the output of another process and that the systems of the network are connected to form a tree topology,
preventing the presence of cycles.\\
In this paper we will formally address this kind of network according to the following definition.
\begin{define}
	Consider the ensemble of a rooted tree topology of $n$ nodes $N_j$ and a
	corresponding set of $n$ linear time-discrete SISO systems
	affected by noise, described according to the model \eqref{eq:noisy sys}.
	Namely, assume $N_i$ as the root node.
	Moreover, let $\{\varrho_j\}_{j=1,\ldots,n}$ be zero-mean wide-sense stationary
	random processes satisfying \eqref{eq:uncorrelation}, i.e. mutually 
	not correlated zero-mean noises.
	Then, we define \textit{Linear Cascade Model Tree	(LCMT)} a dynamical network
	defined by the equation system
	\begin{align}
	\label{eq:LCMT representation}
		\left\{
		\begin{array}{l}
			X_1 = H_1(z)X_{\pi_1} + \varrho_1 \\
			\ldots \\
			X_n = H_n(z)X_{\pi_n} + \varrho_n \,,
		\end{array}
		\right.
	\end{align}
	where $H_i(z)\equiv0$ and the set $\{\pi_1,\ldots,\pi_n\}$ is a permutation
	of $\{1,\ldots,n\}$.
\end{define}
\begin{define}
	A LCMT is \textit{well-posed} if
	$\Phi_{\varrho_j}(\w)>0$ for all $\varrho_j,$ and for all $\w$
\end{define}

Assuming to have a complete statistical knowledge of each process $\{X_i\}_{i=1,\ldots,n}$, we are interested in the identification of the links, which describe the tree characterizing the network topology.
To this aim, hereafter we introduce some preliminary results, which can be exploited to define a mathematical tool for the quantitative characterization of the connections.

Let us consider two stochastic processes $X_i$, $X_j$ and let $W_{ji}(z)$ be a time-discrete SISO transfer function.
Hence, consider the quadratic cost
\begin{align}\label{eq:cost}
	E\left[(\eps_Q)^2 \right] \,,
\end{align}
where
\begin{align*}
	\eps_Q\doteq Q(z)(X_j - W_{ji}(z)X_i)
\end{align*}
and $Q(z)$ is an arbitrary stable and causally invertible
time-discrete transfer function weighting the error 
\begin{align*}
	e_{ji}\doteq X_j - W_{ji}(z)X_i \,.
\end{align*}
Then, the computation of the transfer function $\hat W(z)$ that minimizes the quadratic
cost \eqref{eq:cost} is a well-known problem in scientific literature and its
solution is referred to as the Wiener filter \cite{ksh00}.
\begin{prop}[Wiener filter]
	The Wiener filter modeling $X_j$ by $X_i$ is the linear stable
	filter $\hat W_{ji}$ minimizing the filtered quantity
	\eqref{eq:cost}.
	Its expression is given by
	\begin{align}\label{eq:noncausalwiener}
		\hat W_{ji}(z) = \frac{\Phi_{X_i X_j}(z)}{\Phi_{X_i}(z)}
	\end{align}
	and it does not depend upon $Q(z)$.
	Moreover, the minimized cost is equal to
	\begin{align*}
		&\min E\left[\eps_{Q}^2\right]= \\
		&=\frac{1}{2\pi}\int_{-\pi}^{\pi} |Q(\w)|^2
			\left(\Phi_{X_j}(\omega)
			-|\Phi_{X_j X_i}(\w)|^2 \Phi_{X_i}^{-1}(\w) \right) d\w \,,
	\end{align*}
	and the corresponding error
	\begin{align*}
		\hat{e}_{ji}\doteq X_j - \hat{W}_{ji}(z)X_i
	\end{align*}
	is not correlated with $X_i$, i.e.
	\begin{align}
		E[\hat{e}_{ji}X_i] = 0 \,.
	\end{align}
\end{prop}
\begin{IEEEproof}
See, for example, \cite{ljung99,ksh00}.
\end{IEEEproof}

Since the weighting function $Q(z)$ does not affect the Wiener filter, but only the energy of the filtered error, we can choose $Q(z)$ equal to $F_j(z)$, the inverse of the
spectral factor of $\Phi_{X_j}(z)$, that is
\begin{align}
	\Phi_{X_j}(z)=F_j^{-1}(z)(F_j^{-1}(z))^* \,.
\end{align}
In particular, it is worth recalling that $F_j(z)$ is stable and causally invertible \cite{kail01}.
Therefore, the minimum of cost \eqref{eq:cost} assumes the value
\begin{align}\label{eq:coherentcost}
	\min E[\eps_{F_j}^2] = \frac{1}{2\pi}\int_{-\pi}^{\pi}
			\left(1- \frac{|\Phi_{X_j X_i}(\w)|^2}
				{\Phi_{X_i}(\w)\Phi_{X_j}(\w)}
			\right) d\w \,.
\end{align}
Observe that, due to such choice of $Q(z)$, the cost turns out to explicitly depend
on the \emph{coherence function} of the two processes:
\begin{align}
	C_{X_iX_j}(\w)\doteq\frac{|\Phi_{X_jX_i}(\w)|^2}
					{\Phi_{X_i}(\w)\Phi_{X_j}(\w)} \,.
\end{align}
Let us recall that the coherence function is not negative and symmetric with respect to $\w$.
Moreover, it is also well-known that the cross-spectral density satisfies the
Schwartz inequality and, thus, the coherence function results limited between $0$ and $1$.
Therefore, according to the previous results, the cost \eqref{eq:coherentcost} turns out to be dimensionless and not depending on the ``energy'' of the stochastic processes $X_i$ and $X_j$.

The following result holds.
\begin{prop}\label{pr:trinequality-dynamic}
	In a well-posed LCMT, the binary function
	\begin{align}\label{eq:distance}
		d(X_i,X_j)\doteq
				\left[ \frac{1}{2\pi}
					 \int_{-\pi}^{\pi}
					\left(1- 
						C_{X_{i} X_{j}}(\w)
					\right) d\w 
				\right]^{1/2}
	\end{align}
	is a metric.
\end{prop}
\begin{IEEEproof}
The only non trivial property to be proved is the triangle inequality.
Let $\hat W_{ji}(z)$ be the Wiener filter between $X_i, X_j$ computed according to \eqref{eq:noncausalwiener} and $e_{ji}$ the relative error.
The following relations hold:
\begin{align*}
	&X_3=\hat W_{31}(z) X_1 + e_{31}\\
	&X_3=\hat W_{32}(z) X_2 + e_{32}\\
	&X_2=\hat W_{21}(z) X_1 + e_{21}.
\end{align*}
Since $\hat W_{31}(z)$ is the Wiener filter between the two
processes $X_1$ and $X_3$, it performs better at any frequency
than any other linear filter, such as
$\hat W_{32}(z) \hat W_{21}(z)$.
So we have
\begin{align*}
	\Phi_{e_{31}}(\w)
		&\leq \Phi_{e_{32}}(\w) +
		|\hat W_{32}(\w)|^2 \Phi_{e_{21}}(\w) +\\
		&+\Phi_{e_{32} e_{21}}(\w)  \hat W_{32}^*(\w)+
		\hat W_{32}(\w) \Phi_{e_{21} e_{32}}(\w) \leq\\
		&\leq(\sqrt{\Phi_{e_{32}}(\w)}
			 + |\hat W_{32}(\w)|\sqrt{\Phi_{e_{21}}(\w)})^2
		\quad \forall~\w\in\mathbb{R}.
\end{align*}
For the sake of simplicity we neglect to explicitly write
the argument $\w$ in the following passages. 
Normalizing with respect to $\Phi_{X_{3}}$, we find
\begin{align*}
	\frac{\Phi_{e_{31}}}{\Phi_{X_3}}\leq
		\frac{1}{\Phi_{X_{3}}}
		(\sqrt{\Phi_{e_{32}}} +
			|\hat W_{32}|\sqrt { \Phi_{e_{21}} })^2
\end{align*}
and considering the 2-norm properties
\begin{align*}
	&\left( \int_{-\pi}^{\pi}
		\frac{\Phi_{e_{31}}}{\Phi_{X_3}}d\w\right)^\frac{1}{2} \leq \\
	&\leq\left( \int_{-\pi}^{\pi}
		\frac{\Phi_{e_{32}}}{\Phi_{X_{3}}}d\w
		\right)^\frac{1}{2}
		+\left( \int_{-\pi}^{\pi}
			\frac{|\Phi_{X_3 X_2}|^2}{ \Phi_{X_{3}} \Phi_{X_{2}}}
			\frac{ \Phi_{e_{21}} } {\Phi_{X_{2}}}d\w
		\right)^\frac{1}{2} \,,
\end{align*}
where we have substituted the expression of $\hat W_{32}$.
Finally, observing that
\begin{align*}
	0\leq \frac{|\Phi_{X_3 X_2}|^2}{ \Phi_{X_{3}} \Phi_{X_{2}}}
		\leq 1,
\end{align*}
we find
\begin{align*}
	d(X_1,X_3) \leq d(X_1,X_2) + d(X_2,X_3).
\end{align*}
\end{IEEEproof}

\section{Main result}\label{sec:theory}
In this section we exploit the coherence-based distance \eqref{eq:distance} to derive sufficient conditions to guarantee the exact reconstruction of the topology of a dynamical network.
To this end, we first need to introduce a few definitions and technical lemmas.
\begin{define}
	We define ``path'' from $N_i$ to $N_j$
	a finite sequence of $l>0$ nodes $N_{\pi_1},..., N_{\pi_l}$
	such that
	\begin{itemize}
		\item $N_{\pi_1}=N_i$
		\item $N_{\pi_l}=N_j$
		\item $N_{\pi_i}$ and $N_{\pi_{i+1}}$ are linked by an arc of the tree
			for $i=1,...,l-1$
		\item $N_{\pi_i}\neq N_{\pi_j}$ for $i\neq j$.
	\end{itemize}
\end{define}

In the following we consider LCMT networked systems.
It is worth underlining that a rooted tree is a pair made of a tree and one of its nodes $N_r$, named as ``root''.
Hence, since a tree is a connected graph, in a LCMT network there is always a path between two nodes and, since there are no cycles, such a path is also unique.\\
The presence of a special node labeled as ``root'' induces a natural
relation of ``order'' among the nodes in the following way
\begin{define}
	Given a rooted tree, consider the path from $N_r$ to another node $N_j$.
	A node $N_i$ is said to be an ancestor of $N_j$ if $N_i \neq N_j$
	and if it belongs to the path from $N_r$ to $N_j$.
	Alternatively, we say that $N_j$ is a descendant of $N_i$.
	We also say that $N_i$ is parent of $N_j$ (or that $N_j$ is a child of $N_i$)
	if, in addition, $N_j$ and $N_i$ are connected by an arc.
\end{define}

It is straightforward to prove that the root is an ancestor to all the other nodes and
that every node but the root has exactly one parent.
Hereafter an important result about the correlation property in a LCMT is introduced.
\begin{lem}\label{lem:uncorrelation signal noise}
	Given a LCMT $\mathcal{T}$, consider a node $N_j$ and a node $N_i\neq N_j$
	which is not a descendant of $N_j$.
	Then it holds that $E[\varrho_j X_i]=0$.
\end{lem}
\begin{IEEEproof}
	Let $N_r$ be the root of $\mathcal{T}$ and $N_{\pi_1},...,N_{\pi_l}$
	the path from $N_r$ to $N_i$.
	Exploiting the linear dependencies among the signals of the LCMT,
	$X_i$ can be expressed in terms of the noises $\varrho_{\pi_1},...,\varrho_{\pi_l}$
	\begin{align}
		X_i=\sum_{q=1}^{l} W_{i \pi_q} \varrho_{\pi_q}
	\end{align}
	where
	\begin{align}\label{eq:productTF}
		W_{i \pi_q}=\prod_{h=q}^{l-1} H_{\pi_h}.
	\end{align}
	Since $N_i$ is not a descendant of $N_j$ and $N_i\neq N_j$, we have that
	$\varrho_{\pi_q}\neq \varrho _{j}$ for $q=1,...,l$, thus
	\begin{align}
		E[\varrho_j X_i]=
			E\left[\varrho_j\sum_{q=1}^{l} W_{i \pi_q} \varrho_{\pi_q}\right]=0
	\end{align}
\end{IEEEproof}

The two following lemmas provide two important inequalities about the coherence
functions related to the network signals.
\begin{lem}
\label{lem:cousins}
	Consider a LCMT $\mathcal{T}$ and three nodes $N_i$, $N_j$ and $N_k$ such
	that 
	\begin{itemize}
		\item $N_k$ is a descendant of $N_j$
		\item $N_i$ is not a descendant of $N_j$ and $N_i\neq N_j$.
	\end{itemize}
	Then we have that $C_{X_i X_j} \geq C_{X_i X_k}$.
	Moreover, if $\mathcal{T}$ is well-posed then the inequality is strict.
\end{lem}
\begin{IEEEproof}
	Consider the path from $N_j$ to $N_k$ described by the sequence
	$N_{\pi_1},...,N_{\pi_l}$.
	Exploiting the linear relations (\ref{eq:noisy sys} ),
	the process $X_k$ can be expressed in terms of $X_j$ and of the noises
	acting on the nodes $N_{\pi_2},...,N_{\pi_l}$ which are all descendants of $N_j$.
	\begin{align}
		X_k=W_{k \pi_1}X_j +\sum_{q=2}^{l} W_{k \pi_q} \varrho_{\pi_q}
	\end{align}
	where $W_{i \pi_q}$ is defined as in (\ref{eq:productTF}).
	Now, we intend to evaluate the coherence between $X_i$ and $X_j$.
	From the assumption on $N_i$, it follows that $N_i$ is not on the path
	from $N_j$ to $N_k$. In other words, $N_i$ is not a descendant of $N_{\pi_q}$
	and $N_i \neq N_{\pi_q}$ for $q=1,...,l$.
	We can write
	\begin{align}
		&C_{X_i X_k}=\frac{|\Phi_{X_i X_k}|^2}{\Phi_{X_i}\Phi_{X_k}}= \nonumber\\
			&\quad=\frac{|W_{k \pi_1}|^2|\Phi_{X_i X_j}|^2}
	{\Phi_{X_i}[\Phi_{X_j}|W_{k \pi_1}|^2+\sum_{q=2}^{l}|W_{k \pi_q}|^2\Phi_{\varrho_{\pi_q}}]}
	\end{align}
	where the last equality holds because of Lemma \ref{lem:uncorrelation signal noise} .
	Collecting the factor $\Phi_{X_j}|W_{k \pi_1}|^2$, we obtain
	\begin{align}
		C_{X_i X_k}=\frac{|\Phi_{X_i X_j}|^2}
		{\Phi_{X_i}\Phi_{X_j}\left[1+
		\frac{\sum_{q=2}^{l}|W_{k \pi_q}|^2\Phi_{\varrho_{\pi_q}}}{\Phi_{X_j}|W_{k \pi_1}|^2}\right]}
			\leq C_{X_i X_j}
	\end{align}
	where the inequality is strict if $\sum_{q=2}^{l}|W_{k \pi_q}|^2\Phi_{\varrho_{\pi_q}}>0$.
\end{IEEEproof}
\begin{lem}\label{lem:nephews}
	Consider a LCMT $\mathcal{T}$ and three different
	nodes $N_i$, $N_j$ and $N_k$ such that 
	\begin{itemize}
		\item $N_k$ is a child of $N_j$
		\item $N_i\neq N_j,N_k$ and it is not a descendant of $N_k$
	\end{itemize}
	Then $C_{X_j X_k} \geq C_{X_i X_k}$.
	Moreover, if $\mathcal{T}$ is well-posed the inequality is strict.
\end{lem}
\begin{IEEEproof}
	Assume that $X_k=H_{kj}X_j+\varrho_k$ and let us distinguish two possible scenarios.\\
	\textbf{case A}\\
	First, consider the case where $N_j$ is a descendant of $N_i$.
	Consider the path from $N_i$ to $N_j$ described by the sequence of $l$ nodes
	$N_{\pi_1},...,N_{\pi_l}$ where $N_{\pi_1}=N_{i}$ and $N_{\pi_l}=N_j$.
	The process $X_j$ can be expressed in terms of $X_i$ and of the noises
	acting on the nodes $N_{\pi_2},...,N_{\pi_l}$ which are all descendants of $N_i$.
	\begin{align}
		X_j=W_{j \pi_1}X_i +\sum_{q=2}^{l} W_{j \pi_q} \varrho_{\pi_q} \,.
	\end{align}
	Exploiting Lemma \ref{lem:uncorrelation signal noise} we can evaluate
	the following quantities
	\begin{align}
		&C_{X_i X_k}=\frac{|\Phi_{X_i X_k}|^2}{\Phi_{X_i}\Phi_{X_k}}=
		\frac{
			|W_{j \pi_1}|^2|H_{kj}|^2
		|\Phi_{X_i}|^2}
		{\Phi_{X_i}  \Phi_{X_k}}= \nonumber\\
		&  \qquad \quad =\frac{|W_{j \pi_1}|^2|H_{kj}|^2\Phi_{X_i}}
		{\Phi_{X_k}}
	\end{align}
	and
	\begin{align}
		&C_{X_j X_k}=\frac{|\Phi_{X_j X_k}|^2}{\Phi_{X_j}\Phi_{X_k}}=
		\frac{|H_{kj}|^2|\Phi_{X_j}|^2}
		{\Phi_{X_j}  \Phi_{X_k}}= \nonumber\\
		& \qquad \quad =\frac{|H_{kj}|^2}
		{\Phi_{X_k}}\left[ \Phi_{X_i}|W_{j \pi_1}|^2+\sum_{q=2}^{l}|W_{j \pi_q}|^2\Phi_{\varrho_{\pi_q}} \right]
	\end{align}
	By inspection we have the assertion.\\
	Now we are left to consider the case where $N_j$ is not a descendant
	of $N_i$.
	Then, also  $N_k$ is not a descendant of $N_i$. By hypothesis, $N_i$ is not a descendant of $N_k$, either. Thus, they must have
	a common ancestor $N_d$, such that the two paths from $N_d$ to $N_k$ and from $N_d$ to $N_i$ have only $N_d$ in common.
	Consider the path from $N_d$ to $N_i$, such that it is possible to write
	\begin{equation}
		X_i=W_{i \pi_1}X_d +\sum_{q=2}^{l} W_{i \pi_q} \varrho_{\pi_q}.
	\end{equation}
	Exploiting lemma \ref{lem:uncorrelation signal noise}, we have
	\begin{align}
		& C_{X_i X_k}=\frac{|\Phi_{X_i X_k}|^2}{\Phi_{X_i}\Phi_{X_k}}=\\
		& \quad =\frac{|\Phi_{X_k X_d}|^2}{\Phi_{X_k}\left[\Phi_{X_d}{+\sum_{q=2}^{l}|W_{j \pi_q}|^2\Phi_{\varrho_{\pi_q}}}\right]}\leq \\
		& \qquad \leq C_{X_k X_d}
	\end{align}
	If $N_d=N_j$, we have the assertion. If $N_d\neq N_j$, then $N_j$ must be a descendant of $N_d$. We are in a situation equivalent to case A: there is a node $N_d$ such that $N_j$ is one of its descendants. As a consequence, we can state that
	\begin{equation}
		C_{X_k X_d}\leq C_{X_k X_j }.
	\end{equation}
	Combining the last two inequalities, we conclude that the lemma holds also in this case.
\end{IEEEproof}

All the previous lemmas are functional to the show that the coherence distance
(\ref{eq:distance}) is minimal between two contiguous nodes, as summarized in this theorem.
\begin{theo}\label{th:incident node theorem}
	Given a LCMT $\mathcal{T}$, consider a node $N_a$ and a node $N_b\neq N_a$
	which is not directly linked to it.
	Then there exists a node $N_c$ directly linked to $N_a$ such that
	\begin{align}
		d(N_a, N_c)\leq d(N_a, N_b)
	\end{align}
	where the inequality is strict if $\mathcal{T}$ is well-posed.
\end{theo}
\begin{IEEEproof}
	First, consider the case where $N_b$ is a descendant of $N_a$.
	Name $N_c$ the child of $N_a$ on the path linking it to $N_b$.
	Since $N_c$ is directly linked to $N_a$, we have $N_b\neq N_c$.
	Moreover $N_b$ is a descendant of $N_c$.
	We are allowed to apply lemma (\ref{lem:cousins})
	with $N_i=N_a$, $N_j=N_c$ and $N_k=N_b$
	to have the assertion.\\
	Now, consider the case where $N_b$ is not a descendant of $N_a$.
	$N_a$ can not be the root, otherwise $N_b$ would be one of its descendants.
	Thus $N_a$ has a parent and let us name it $N_c$.
	$N_b$ can not be $N_c$ because it is not directly linked to $N_a$.
	Applying lemma (\ref{lem:nephews}) with $N_i=N_b$, $N_j=N_c$ and $N_k=N_a$
	and by the definition of the coherence distance (\ref{eq:distance}),
	we have the assertion.\\
\end{IEEEproof}

Theorem \ref{th:incident node theorem} can be fruitfully exploited to determine whether two processes in a well-posed LCMT are directly linked.
Nonetheless, when we are dealing with data sampled from actual systems the computation of $d$, that is of the coherence function, is affected by the limited time horizon of the observations.
However, the estimates of the spectral and cross-spectral densities converge to the actual values as the time horizon approaches infinity.
Hence, in the following we will assume to sample the processes over a sufficiently large time interval.

We are ready to show the main contribution of the paper.
\begin{theo}
	Consider a well-posed LCMT $\mathcal{T}$ and assume to observe the signals
	$X_j$ during a time horizon $t$.
	Compute an estimate of the coherence based distances $d_{ij}=d(X_i, X_j)$ among the nodes $N_j$	and evaluate the relative Minimum Spanning Tree (MST). When $t$ approaches infinity, the corresponding topology is equivalent to the unique MST $T$ associated to the coherence metric.
\end{theo}
\begin{IEEEproof}
	The proof consists in showing that the MST $T$ associated to the distance (\ref{eq:distance}) is unique and corresponds to the LCMT topology.
	We will prove this result by induction on the number $n$ of nodes of the LCMT.\\
	The basic induction step consists in observing that theorem is true for $n=2$.\\
	Now assume the theorem true for a LCMT with $n$ nodes.
	Given a LCMT $\mathcal{T}$ with $n+1$ nodes, remove one of its ``leaves''. By leaf we mean a non-root node with no descendants. This operation is always possible since any rooted tree with at least two nodes has at least one leaf. Without loss of generality, let the removed leaf be $N_{n+1}$ and let $N_{i}$ be its parent.
	Now we have a LCMT $\mathcal{T'}$ with $n$ nodes and with the same topology of $\mathcal{T}$ apart from the removed arc $(i,n+1)$.
	Using the induction hypothesis, we know that the topology of $\mathcal{T'}$ is given by the unique MST $T'$ obtained considering the distances among the nodes $N_1, ..., N_n$. Now compute
	\begin{equation}\label{eq:minimization to reconstruct}
		i^* = \arg\min_{j<N+1} d(X_i,X_{n+1}).
	\end{equation}
	The solution of such a minimization problem is unique since the LCMT $\mathcal{T}$ is well posed. Because of lemma \ref{th:incident node theorem}, the arc $(i^*,N+1)$ belongs to the topology of $\mathcal{T}$, so we conclude $i^*=i$.
	Let $T$ be the spanning tree obtained by adding the arc $(i,N+1)$ to $T'$. So far, we have shown that $T$ represents the topology of $\mathcal{T}$. We have to prove that $T$ is the unique MST related to the distance (\ref{eq:distance}) among the nodes $N_1, ..., N_{n+1}$.
	Suppose, by contradiction, that there is a minimum spanning tree $\bar T\neq T$ with weight lesser or equal than the weight of $T$. The only arc of $\bar T$ incident to the node $N_{n+1}$ is $(i,n+1)$. If there were another arc $(k,n+1)$ in $\bar T$ we could replace it with the arc $(k,i)$ obtaining a spanning tree with inferior cost.
	Indeed, by lemma \ref{lem:nephews}, we would have
	\begin{equation}
		d(X_k,X_i)<d(X_{n+1},X_i).
	\end{equation}
	So, if $\bar T$ is a minimum spanning tree, then $X_{n+1}$ can be connected only to $X_i$. Let $\bar T'$ be the tree obtained by $\bar T$ removing the arc $(i,n+1)$. $\bar T'$ is the minimum spanning tree for the nodes $N_1, ..., N_{n}$ since it has been obtained from $\bar T$ removing the node $N_{n+1}$ which has a single connection. However, by the induction hypothesis, there is a unique MST $T'$ among the nodes $N_1, ..., N_{n}$. Thus we have that $\bar T'=T'$. It immediately follows the contradiction that $\bar T=T$. 
\end{IEEEproof}

So far, we have assumed that the dynamics of the network is described
by a rooted tree. Moreover, the previous theorem proves that the topology
structure can be correctly identified evaluating the MST according
to the distance (\ref{eq:distance}).
However, no information is recovered about the root node.
The following result shows that such an information is not necessary (or, equivalently, not recoverable). Indeed, from a modeling point of view, the choice of the root
can be arbitrary (as long as we are considering non-causal transfer functions linking
the processes $X_j$).
\begin{theo}\label{th:equivalent LCMTs}
	Given a LCMT $\mathcal{T}$ whose root is the node $N_j$ and given one of its
	children $N_i$, it is possible to define another LCMT $\mathcal{T}^*$
	with the same tree structure and described by the same
	processes $X_k$, $k=1,...,n$, such that its root is $N_i$.
\end{theo}
\begin{IEEEproof}
	Consider the Wiener Filter $W_{ji}$ modeling the signal $X_j$, seen as the output,
	when $X_i$ is the input
	\begin{equation}
		X_j=W_{ji} X_i + e_{ji}.
	\end{equation}
	Now, consider a rooted tree with the same topology of $\mathcal{T}$ but with
	$N_i$ as the root.
	Define $H_k^*=H_k$ and $\varrho^*_k=\varrho_k$
	for all $k\neq i,j$. Conversely, define
	\begin{align}
		& H^*_j=W_{ji}
		& \varrho^*_j=e_{ji}\\
		& H^*_i\equiv0
		& \varrho^*_i=X_i.
	\end{align}
	To show that the new dynamical network with $N_i$ as root and described
	by the filters $H^*_k$ is an LCMT,
	we need to prove that, for $h \neq k$,
	\begin{align}
		E[\varrho^*_h \varrho^*_k]=0.
	\end{align}
	There are three possible scenarios.\\
	If $h=i$ and $k=j$ or $h=j$ and $k=i$, then
	\begin{align}
		E[\varrho^*_h \varrho^*_k]=0.
	\end{align}
	because of the Wiener Filter properties.\\
	If $h=i,j$ and $k\neq i,j$ (or equivalently $h\neq i,j$ and $k=i,j$),
	then lemma \ref{lem:uncorrelation signal noise} can be applied.\\
	If $h\neq i,j$ and $k\neq i,j$, then
	\begin{align}
		E[\varrho^*_h \varrho^*_k]=E[\varrho_h \varrho_k]
	\end{align}
	and we have the assertion because $\varrho_h$ and $\varrho_k$ are two
	noise signals of the original LCMT $\mathcal{T}$.
\end{IEEEproof}

It is straightforward to show that, starting from an LCMT $\mathcal{T}$, we can arbitrary define a LCMT $\mathcal{T^*}$ having an arbitrary node as root. Indeed, it is sufficient to iteratively apply Theorem \ref{th:equivalent LCMTs} along the path starting from the original root to the new one.

\section{Numerical examples}\label{sec:numeric}
In this section we introduce a suitable framework to illustrate the application of the previous theoretical results to numerical analysis.
It is worth observing that the previous results have been developed for the most general class of linear models.
Indeed, no assumptions have been done on the order and causality property of the considered transfer functions.\\
Moreover, let us highlight that the coherence based analysis must be realized ``off-line'', since the processes have to be evaluated over their entire time span.
Thus, because the coherence function can be numerically computed only over limited intervals, in the following examples we will consider sufficiently long time spans to reduce the numerical error.

Hence, let us build the original dynamical networks according the following rules:
\begin{itemize}
	\item each system is described according to the model \eqref{eq:noisy sys};
	\item each transfer function $H_j$ is randomly generated and such that it is causal and at most of the 		second order;
	\item the tree topology is randomly chosen;
	\item the noises $\varrho_j$ are numerically generated with a pseudo-random algorithm;
	\item the noise-to-signal ratio of each system is equal to one.
\end{itemize}
Then, such networks are simulated over 1000 time steps and the related data $X_j$ are collected.
The corresponding coherence based distances are evaluated and used for the extraction of the MST, that defines the link topology.

\begin{figure}
	\begin{center}
		\includegraphics[width=0.98\columnwidth]{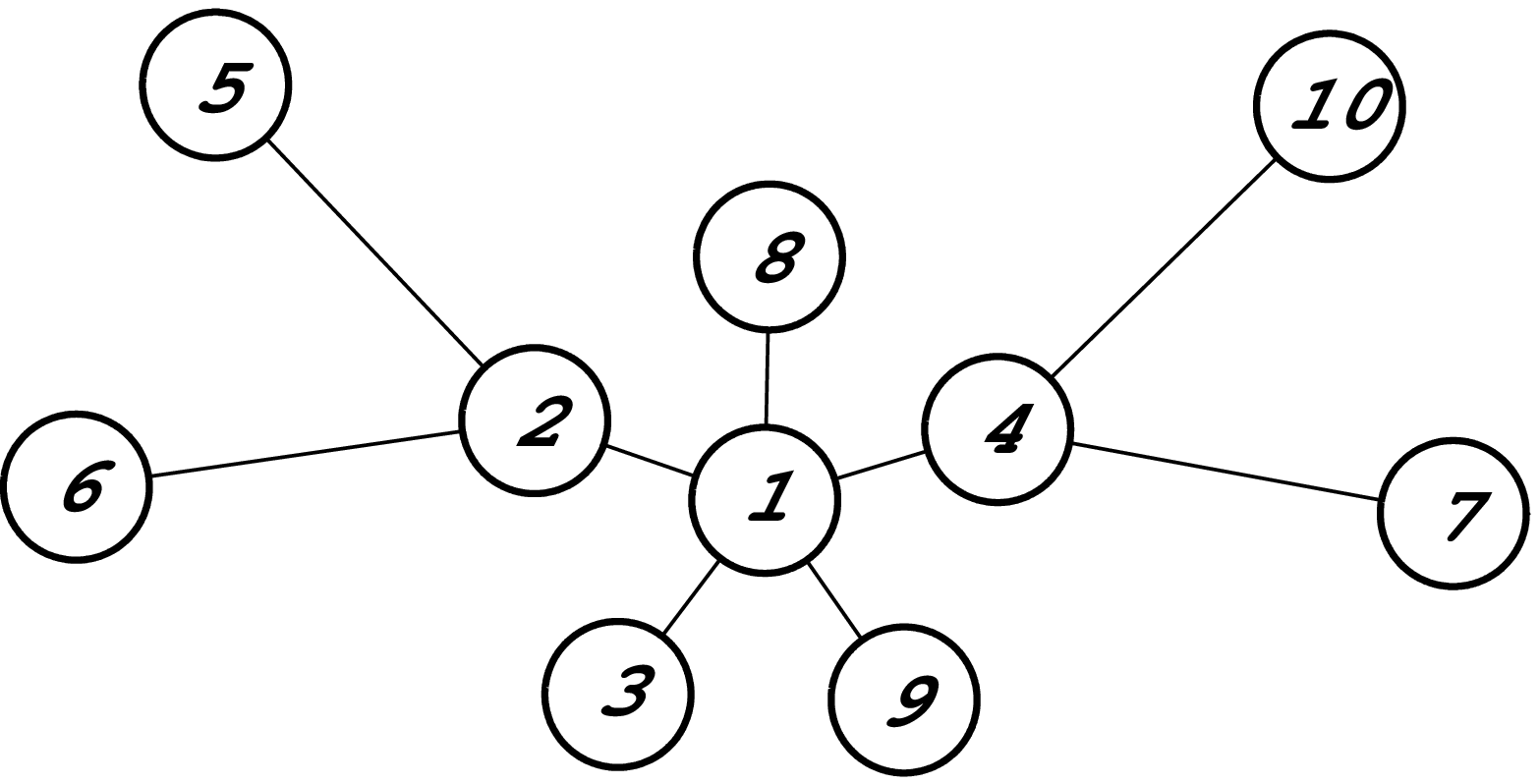}
	\end{center}
	\caption{The figure illustrates the topology of the 10 nodes network analyzed
		in the numerical examples paragraph.
		Each node is responsible for a process $X_j$, while the arcs describe the
		connections among them, according to the linear SISO model \eqref{eq:noisy sys}.
		For the data generation we have considered only transfer functions of
		at most the second order.
		The noises $\varrho_j$ have been assumed to provide half the power of the
		affected processes.
		The samples have been collected over 1000 time steps.}
		\label{fig:truetreeN10_002}
\end{figure}

The above procedure will be first applied to a ten node network.
In particular, to test the numerical reliability of the topological identification technique, we repeat such analysis several times, so that a significant number of network configurations is considered.
The corresponding results fit the expectations and the real topology is correctly identified each time.
In Fig. \ref{fig:truetreeN10_002} one of the considered network configurations is depicted, while the related coherence based distance matrix is reported in Table \ref{tab:coherencedistsim}.
\begin{table*}
	{
	\centering
	\begin{tabular}{|c|cccccccccc|}
	\hline
		& $X_1$ & $X_2$ & $X_3$ & $X_4$ & $X_5$ & $X_6$ & $X_7$ & $X_8$ & $X_9$ & $X_{10}$ \\
		\hline
		$X_1$ & 0	& 0,7299	& 0,6675	& 0,7351	& 0,8316	& 0,8542	& 0,8297	& 0,7055	& 0,6549	& 0,8298 \\
		$X_2$ & 0,7299	& 0	& 0,8065	& 0,8353	& 0,6934	& 0,7358	& 0,8786	& 0,8483	& 0,8299	& 0,8717 \\
		$X_3$ & 0,6675	& 0,8065	& 0	& 0,8216	& 0,8744	& 0,8807	& 0,8750	& 0,8262	& 0,7841	& 0,8821 \\
		$X_4$ & 0,7351	& 0,8353	& 0,8216	& 0	& 0,8662	& 0,8722	& 0,7404	& 0,8502	& 0,8198	& 0,7039 \\
		$X_5$ & 0,8316	& 0,6934	& 0,8744	& 0,8662	& 0	& 0,8540	& 0,8919	& 0,8995	& 0,8730	& 0,8846 \\
		$X_6$ & 0,8542	& 0,7358	& 0,8807	& 0,8722	& 0,8540	& 0	& 0,8934	& 0,8984	& 0,8796	& 0,8944 \\
		$X_7$ & 0,8297	& 0,8786	& 0,8750	& 0,7404	& 0,8919	& 0,8934	& 0	& 0,8838	& 0,8694	& 0,8346 \\
		$X_8$ & 0,7055	& 0,8483	& 0,8262	& 0,8502	& 0,8995	& 0,8984	& 0,8838	& 0	& 0,8167	& 0,8908 \\
		$X_9$ & 0,6549	& 0,8299	& 0,7841	& 0,8198	& 0,8730	& 0,8796	& 0,8694	& 0,8167	& 0	& 0,8715 \\
		$X_{10}$ & 0,8298	& 0,8717	& 0,8821	& 0,7039	& 0,8846	& 0,8944	& 0,8346	& 0,8908	& 0,8715	& 0 \\
		\hline
	\end{tabular}
	\caption{the coherence based distance matrix associated to the network topology depicted in Fig. \ref{fig:truetreeN10_002}}
	\label{tab:coherencedistsim}
	}
\end{table*}

\begin{figure}
	\begin{center}
		\includegraphics[width=0.98\columnwidth]{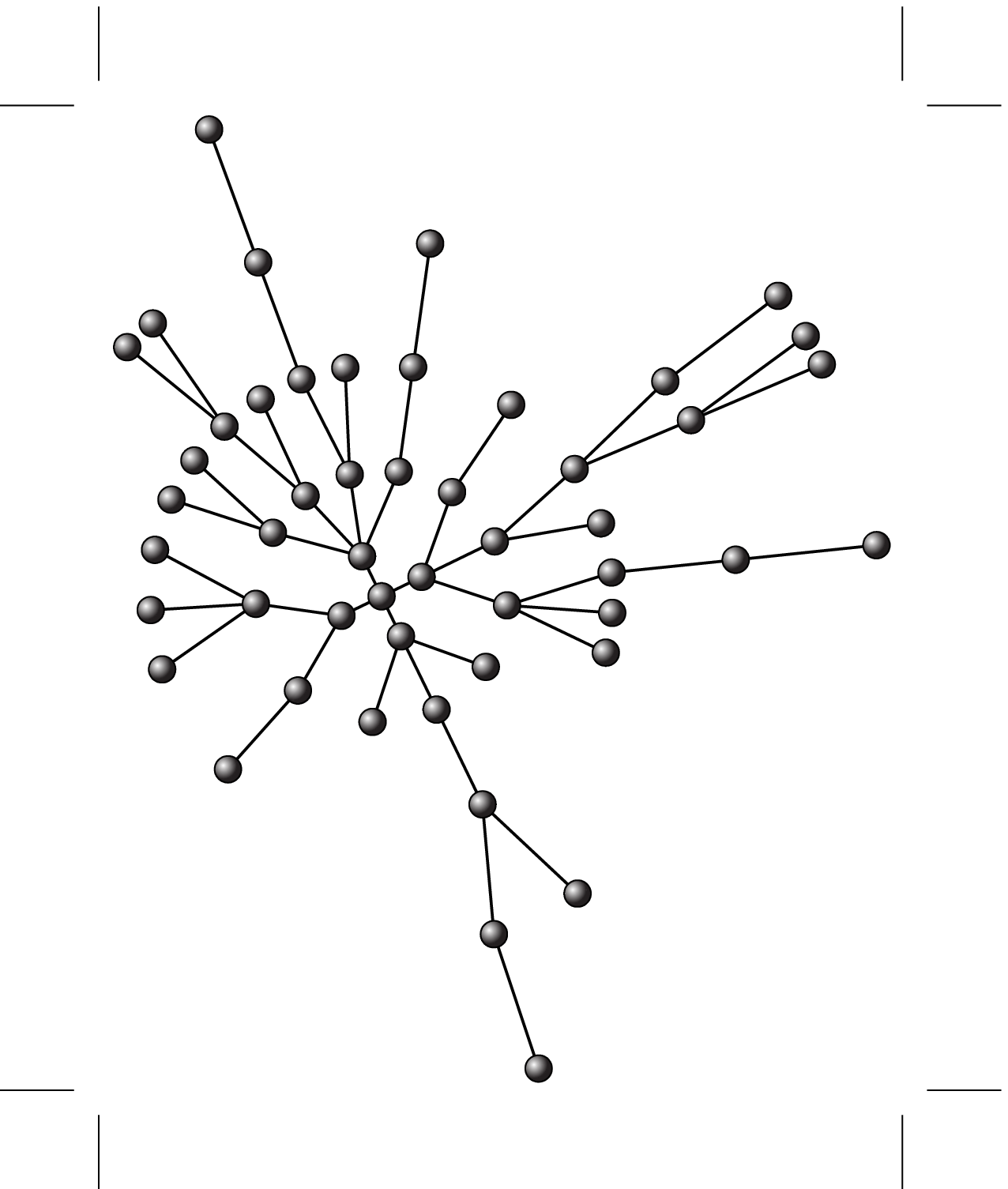}
	\end{center}
	\caption{
		A representative topological configuration of the 50 nodes network case
		considered of the numerical examples paragraph.
		The example has been designed according to the same assumptions of the
		ten node network of Figure \ref{fig:truetreeN10_002}.}
		\label{fig:topology50processes}
\end{figure}

To provide a further test, a new set of similar simulations is performed with a network of fifty dynamical systems, under the same assumptions used in the previous case.
Figure \ref{fig:topology50processes} presents one of the considered network configurations.
For a space limitation issue, we do not report in this manuscript the corresponding coherence based distance matrix.
Nonetheless, the computation of the related MST has successfully identified the real network topology in any of the performed simulations.

\section{Stock market analysis}
\label{sec:portfolio}
In the previous section we have illustrated how the distance \eqref{eq:distance} can be successfully exploited to derive the exact topology of a tree network of linear systems affected by additive noises.
Nonetheless, since the above identification technique is able to catch the most important linear dependencies with respect to the modeling error \eqref{eq:cost}, in the following we present the results obtained by the application of the previous method to the stock market, that is a network of nonlinear systems characterized by multiple dependencies.

\begin{figure*}
	\centering
	\includegraphics[width=0.9\textwidth]{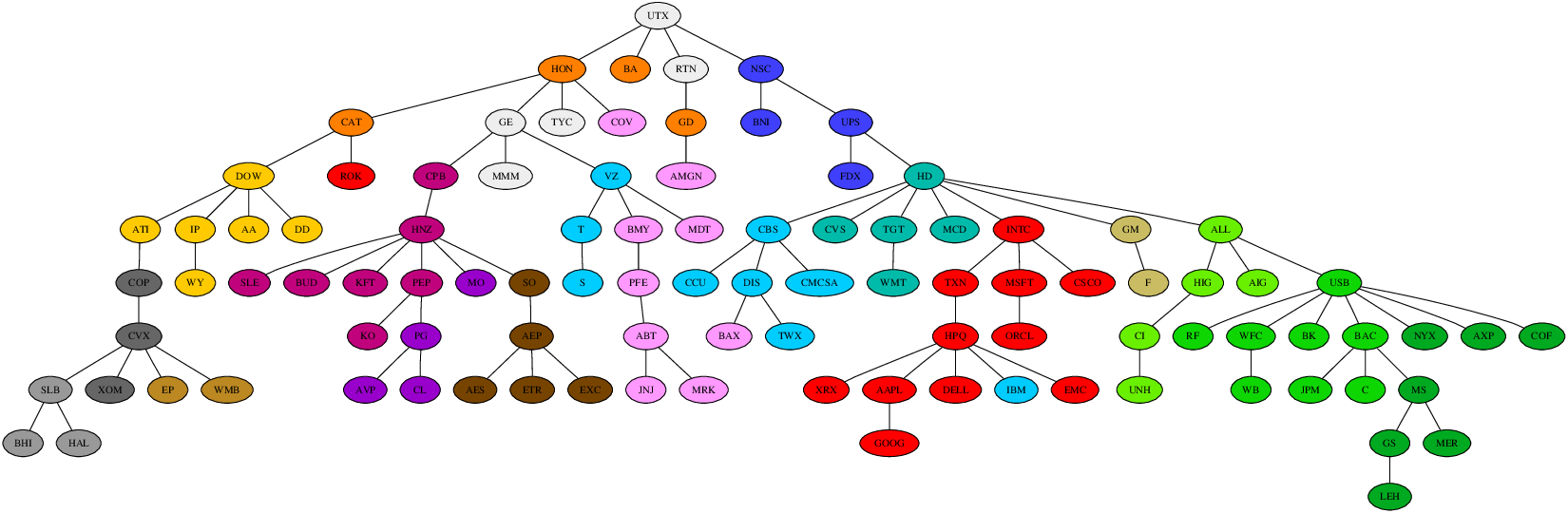}
	\caption{
		(color on line)
		The tree structure obtained using the proposed identification technique.
		Every node represents a stock and the color represents the business sector it belongs to.
		The considered sectors are
		\texttt{Basic Material} (yellow),
		\texttt{Conglomerates} (white),
		\texttt{Healthcare} (pink),	
		\texttt{Transportations} (dark blue),
		\texttt{Technology} (red),
		\texttt{Capital Goods} (orange),
		\texttt{Utilities} (brown tints),
		\texttt{Consumer} (violet tints),
		\texttt{Financial} (green tints),
		\texttt{Energy} (gray tints)
		\texttt{Services} (light blue tints).
		Using the industry classification given by Google, the \texttt{Financial} sector has also been differentiated among
		Insurance Companies (light green),
		Banks (average green) and
		Investment Companies (dark green);
		\texttt{Services} have been divided in
		Information Technology (cyan) and
		Retail (aquamarina), 
		\texttt{Consumer} in
		Food (plum) and
		Personal-care (purple);
		\texttt{Energy} in
		Oil \& Gas (dark gray) and
		Well Equipment (light gray);
		Utilities in Electrical (dark brown) and Natural Gas (light brown).
		\label{fig:tree_mant_dg} }
\end{figure*}

Financial systems are, in general, very complex and deriving information from stock markets is a formidable and challenging task, indeed.
Moreover, it might seem very reductive the attempt to describe the dependencies among the price trends in terms of linear SISO systems with a tree topology. In fact, we should definitely expect multiple input influences, nonlinear relations and feedbacks. However, we can think of adopting a LCMT in order to detect what are the ``strongest'' links in the network. As noted in \cite{mant99}, such an information could be usefully exploited to check if a given portfolio is balanced or not.
In the following, we report the results obtained by the application of our identification technique.\\
A collection of $100$ stocks of the New York Stock Exchange has been observed for four weeks (twenty market days),  in the lapse 03/03/2008 - 03/28/2008 sampling their prices every $2$ minutes.
The stocks have been chosen on the first $100$ stocks with highest trading volume according to the Standard \& Poor Index at the first day of observation
and they are reported in Table \ref{tab:listSP100}.
\begin{table}[!htb]
	\centering		
	{\tiny
	\begin{tabular}{|c|c|c|}
			\hline
			Name & Code & Sector \\
			\hline \hline
			3M Company				& MMM 	& Conglomerates \\ \hline 
Abbott LAboratories 			& ABT 	& Healthcare \\ \hline 
Aes Corporation				& AES 	& Utilities \\ \hline 
Alcoa Inc.				& AA 	& Basic Materials \\ \hline 
Allegheny Technologies Inc. 		& ATI 	& Basic Materials \\ \hline 
Allstate Corporation 			& ALL	& Financial \\ \hline 
Altria Group 				& MO 	& Consumer/Non-Cyclical \\ \hline 
American Electric Power			& AEP 	& Utilities \\ \hline 
American Express 			& AXP 	& Financial \\ \hline 
American International Group 		& AIG 	& Financial \\ \hline 
Amgen Inc. 				& AMGN 	& Healthcare \\ \hline 
Anheuser Busch				& BUD 	& Consumer/Non-Cyclical \\ \hline 
Apple Inc. 				& AAPL 	& Technology \\ \hline 
AT\&T 					& T 	& Services \\ \hline 
Avon Products 				& AVP	& Consumer/Non-Cyclical \\ \hline 
Baker Hughes Inc.			& BHI	& Energy \\ \hline 
Bank of America 			& BAC	& Financial \\ \hline 
Bank of New York Mellon			& BK	& Financial \\ \hline 
Baxter International 			& BAX 	& Healthcare \\ \hline 
Boeing					& BA 	& Capital Goods \\ \hline 
Bristol Myers Squibb			& BMY 	& Healthcare \\ \hline 
Burlington Northern Santa Fe		& BNI 	& Transportation \\ \hline 
Campbell Soup				& CPB 	& Consumer/Non-Cyclical \\ \hline 
Capital One Financial			& COF 	& Financial \\ \hline 
Caterpillar Inc. 			& CAT 	& Capital Goods \\ \hline 
CBS					& CBS 	& Services \\ \hline 
Chevron					& CVX 	& Energy \\ \hline 
CIGNA 					& CI 	& Financial \\ \hline 
Cisco Systems 				& CSCO 	& Technology \\ \hline 
Citigroup Inc				& C 	& Financial \\ \hline 
Clear Channel Communications		& CCU 	& Services \\ \hline 
Coca-Cola				& KO 	& Consumer/Non-Cyclical \\ \hline 
Colgate Palmolive			& CL 	& Consumer/Non-Cyclical \\ \hline 
Comcast					& CMCSA & Services \\ \hline 
Conoco Phillips 			& COP 	& Energy \\ \hline 
Covidien 				& COV 	& Healthcare \\ \hline 
CVS Caremark				& CVS & Services \\ \hline 
Dell Inc				& DELL 	& Technology \\ \hline 
Dow Chemical Company 			& DOW 	& Basic Materials \\ \hline 
E.I. du Pont de Nemours 		& DD 	& Basic Materials \\ \hline 
El Paso					& EP 	& Utilities \\ \hline 
EMC					& EMC 	& Technology \\ \hline 
Entergy 				& ETR 	& Utilities \\ \hline 
Exelon					& EXC 	& Utilities \\ \hline 
Exxon Mobil 				& XOM 	& Energy \\ \hline 
FedEx					& FDX 	& Transportation \\ \hline 
Ford Motor				& F 	& Consumer Cyclical \\ \hline 
General Dynamics 			& GD 	& Capital Goods \\ \hline 
General Electric 			& GE	& Conglomerates \\ \hline 
General Motors				& GM 	& Consumer Cyclical \\ \hline 
Goldman Sachs Group			& GS 	& Financial \\ \hline 
Google Inc.				& GOOG 	& Technology \\ \hline 
Halliburton				& HAL 	& Energy \\ \hline 
Hartford Financial Services		& HIG 	& Financial \\ \hline 
H. J. Heinz 				& HNZ 	& Consumer/Non-Cyclical \\ \hline 
Hewlett-Packard 			& HPQ 	& Technology \\ \hline 
Home Depot				& HD 	& Services \\ \hline 
Honeywell International			& HON 	& Capital Goods \\ \hline 
Intel					& INTC 	& Technology \\ \hline 
International Business Machines		& IBM 	& Services \\ \hline 
International Paper			& IP 	& Basic Materials \\ \hline 
Johnson \& Johnson			& JNJ 	& Healthcare \\ \hline 
JPMorgan Chase				& JPM 	& Financial \\ \hline 
Kraft Foods				& KFT 	& Consumer/Non-Cyclical \\ \hline 
Lehman Brothers Holding			& LEH 	& Financial \\ \hline 
McDonald's				& MCD 	& Services \\ \hline 
Medtronic				& MDT 	& Healthcare \\ \hline 
Merck					& MRK 	& Healthcare \\ \hline 
Merril Lynch				& MER 	& Financial \\ \hline 
Microsoft				& MSFT 	& Technology \\ \hline 
Morgan Stanley				& MS 	& Financial \\ \hline 
Norfolk Souther Group			& NSC 	& Transportation \\ \hline 
NYSE Euronext				& NYX 	& Financial \\ \hline 
Oracle					& ORCL 	& Technology \\ \hline 
Pespi					& PEP 	& Consumer/Non-Cyclical \\ \hline 
Pfizer Inc.				& PFE 	& Healthcare \\ \hline 
Procter \& Gamble			& PG 	& Consumer/Non-Cyclical \\ \hline 
Raytheon				& RTN 	& Conglomerates \\ \hline 
Regions Financial 			& RF 	& Financial \\ \hline 
Rockwell Automation			& ROK 	& Technology \\ \hline 
Sara Lee				& SLE 	& Consumer/Non-Cyclical \\ \hline 
Schlumberger Limited			& SLB 	& Energy \\ \hline 
Southern				& SO 	& Utilities \\ \hline 
Sprint Nextel				& S 	& Services \\ \hline 
Target 					& TGT 	& Services \\ \hline 
Texas Instruments Inc.			& TXN 	& Technology \\ \hline 
Time Warner				& TWX 	& Services \\ \hline 
Tyco International			& TYC	& Conglomerates \\ \hline 
U. S. Bancorp				& USB 	& Financial \\ \hline 
United Parcel Service			& UPS	& Transportation \\ \hline 
United Technologies			& UTX	& Conglomerates \\ \hline 
UnitedHealth Group Inc.			& UNH 	& Financial \\ \hline 
Verizon Communications 			& VZ 	& Services \\ \hline 
Wachovia				& WB 	& Financial \\ \hline 
Wal-Mart Stores				& WMT 	& Services \\ \hline 
Walt Disney				& DIS 	& Services \\ \hline 
Wells Fargo				& WFC 	& Financial \\ \hline 
Weyerhaeuser Company			& WY	& Basic Materials \\ \hline 
Williams Companies			& WMB 	& Utilities \\ \hline 
	Xerox					& XRX 	& Technology \\ \hline
	\end{tabular}
	\caption{List of the companies considered in the analysis \label{tab:listSP100}}
	}
\end{table}
An a-priori organization of the companies has been assumed in accordance with the sector and industry group classification provided by Google Finance$^{\text{\textregistered}}$, that is also the source of our data.
The whole observation horizon spans almost the whole month of March.
Hence, the corresponding price series can not be considered stationary and the statistical tools can not be successfully employed to analyze the raw data.\\
In literature a variety of techniques for the suppression of trends and periodic components in non-stationary time series exists.
However, we want to stress that the application of such procedures introduces an additional prefiltering phase, which is responsible for the computational burden increase.
Moreover, due to the pre- and post-market sessions, there is a discontinuity between the end value of a day and the opening price of the next one.
We have avoided those problems observing that the observation horizon is naturally divided into subperiods, namely weeks and days.
In addition, a single market session can be considered a time period sufficiently short to assume that the influence of trends and seasonal factors are negligible.
Thus, in our analysis, we have followed the natural approach of dividing the historical
series into twenty subperiods corresponding to single days.
Then, we considered the sessions separately, i.e. we have computed the coherence-based distances \eqref{eq:distance} among the stocks for every single day.
Finally, we have averaged such daily distances over the whole observation horizon and the related results have been exploited to extract the MST, providing the corresponding market structure.\\
We find useful to remark that the computation of the distances for smaller data sets is also better performing and that the averaging procedure provides the desired rejection of trends and seasonal components.
Notably, a similar idea, even if more sophisticated, is at the basis of the method developed in \cite{pos08} to detrend non-stationary time series.\\
The final topology is shown in Figure \ref{fig:tree_mant_dg}.
Every node represents a stock and the color represents the business sector or industry it belongs to.
We note that the stocks are very satisfactorily grouped according to their business sectors.
We stress that the a-priori classification in sectors is not a hard fact by itself and we are not trying to match it exactly.
A company could well be categorized in a sector because of its business, but, at the same time, could show a behaviour similar to and explainable through the dynamics of other sectors.
Actually, we would be very interested into finding results of this kind.
Indeed, in those very cases, our quantitative analysis would provide the greatest contributions detecting in an objective way something which is ``counter-intuitive''.
Thus, we just use such a-priori classification as a tool to check if the final topology makes sense and if, at a general level, our approach provides useful results.
Despite this disclaim, it is worth noting that the \texttt{Financial} (green tints), \texttt{Consumer} (violet tints), \texttt{Basic Materials} (yellow), \texttt{Energy} (gray tints) and \texttt{Transportation} (dark blue) sectors are all perfectly grouped, with no exceptions.
In Fig.~\ref{fig:tree_mant_dg}, we note a subclusterization of the \texttt{Financial} sector, as well.
The \texttt{Consumer} sector shows another prominent subclusterization in the \texttt{Food} (plum) and \texttt{Personal/Healthcare} (purple) industries, while
the \texttt{Energy} sector presents an evident subclusterization into the \texttt{Oil \& Gas} (dark gray) and \texttt{Oil Well Equipment} (light gray).
The \texttt{Utilities/Electricity} companies (dark brown) are, interestingly, a different group.
We also observe a big cluster of companies classified as \texttt{Services} (light blue tints).
We have differentiated them in the two industries \texttt{Retail} and \texttt{Information Technology} using two slightly different colors, respectively aquamarine and cyan.
We also note the presence of three \texttt{Services} companies which are isolated from the other ones: \texttt{V} [Verizon], \texttt{T} [AT\&T], and \texttt{S} [Sprint].
All of them are telephone companies.
This might suggest that this industry should show at least a slightly different dynamics from the other service companies.
Note also how the \texttt{Technology} sector (red) is almost perfectly grouped and how \texttt{IBM}, an IT company, even though classified as a \texttt{Services} company, is located in it.
Finally, the two only automobile companies \texttt{GM} and \texttt{F} [Ford] happen to be linked together.
The analysis of this four weeks of the month of March cleanly shows a taxonomic arrangement of the stocks even though the choice of a tree structure might have seemed quite reductive at first thought.

\section{Conclusions}\label{sec:conclusions}
This work has illustrated a simple but effective procedure to identify 
the structure of a network of linear dynamical systems when the topology
is described by a tree.
To the best knowledge of the authors, the problem of identifying a network
has not yet been tackled in scientific literature.
The approach followed in this paper is based on the definition of a
distance function in order to evaluate if there exists a direct link
between two nodes. A few theoretical results are provided, in particular
to guarantee the correctness of the identification procedure.
An application of the technique to real data has also shown that a tree topology can be sufficient to capture information even in complex situations such as financial stock prices.

\section*{Acknowledgments}
The authors would like to thank Prof. Tim Sauer for his precious suggestions and advices.

\bibliography{tac2008}

\end{document}